\documentclass{elsart}
\usepackage{graphicx}
\begin{document}
\begin{frontmatter}
\title{Explicit solutions for relativistic acceleration and rotation}
\author{Yaakov Friedman }
\address{Jerusalem College of
Technology, Jerusalem 91160 Israel\\email: friedman@jct.ac.il}
\begin{abstract}
The Lorentz transformations are represented by Einstein velocity
addition on the ball of relativistically admissible velocities.
This representation is by projective maps. The Lie algebra of this
representation defines the relativistic dynamic equation. If we
introduce a new dynamic variable, called symmetric velocity, the
above representation becomes a representation by conformal,
instead of projective maps. In this variable, the relativistic
dynamic equation for systems with an invariant plane, becomes a
non-linear analytic equation in one complex variable. We obtain
explicit solutions for the motion of a charge in uniform, mutually
perpendicular electric and magnetic fields. By assuming the Clock
Hypothesis and using these solutions, we are able to describe the
space-time transformations between two uniformly accelerated and
rotating systems.


\end{abstract}

\end{frontmatter}

 \section{Representation of the Lorentz transformations on the ball of relativistically admissible velocities}

 The usual Lorentz space-time
 transformations between two inertial systems $K$ and $K'$, moving with
 relative velocity (boost) $\mathbf{b}$. The space
 axes are chosen to be parallel.
 The  Lorentz transformation $L$ transforms the  space-time coordinates $(t,\mathbf{r})$
 in $K$ of an event to the space-time coordinates $(t',\mathbf{r}')$
 in $K'$ of the same event.
 If we assume that the interval $ds^2=(cdt)^2-d\mathbf{r}^2$ is conserved,
 the resulting space-time transformation  between
systems is called the \index{transformation!Lorentz}
\index{Lorentz!transformation}\emph{Lorentz transformation}. In
the case $\mathbf{b}=(b,0,0)$ the Lorentz transformation $L_v$
from system $K'$ to system $K$ is
\begin{equation}\label{lorentz1}
 \begin{array}{cl}
    t &=\gamma(t'+\frac{bx'}{c^2}) \\
    x & =\gamma(bt'+x') \\
    y & =y' \\
    z & =z',
  \end{array}
\end{equation}
with $\gamma=\frac{1}{\sqrt{1-b^2/c^2}}.$

 This Lorentz transformation for arbitrary relative velocity $\mathbf{b}$ can be rewritten in
the vector and block-matrix notation as:
 \begin{equation}\label{lorentzvect0}  \left(
\begin{array}{c} t\\ \mathbf{r}\end{array} \right)
          =L_{\mathbf{b}}\left( \begin{array}{c}  t'\\ \mathbf{r}' \end{array} \right)=
           \left(
         \begin{array}{cc}
              \gamma \quad& \gamma c^{-2}\mathbf{b}^T \\
              \gamma\mathbf{b}  & \gamma P_{\mathbf{b}}+(I-P_{\mathbf{b}})
          \end{array} \right)
         \left( \begin{array}{c}  t'\\ \mathbf{r}'
          \end{array} \right)
\end{equation}   or
         \begin{equation}\label{lorentzvect}
          L_{\mathbf{b}}\left( \begin{array}{c}  t'\\ \mathbf{r}' \end{array} \right)=\gamma \left(
         \begin{array}{cc}
              1 & c^{-2}\mathbf{b}^T \\
              \mathbf{b} \quad &  P_{\mathbf{b}}+\alpha(I-P_{\mathbf{b}})
          \end{array} \right)
         \left( \begin{array}{c}  t'\\ \mathbf{r}'
          \end{array} \right)
,\end{equation}
 where
\begin{equation}\label{alpha}
  \alpha=\gamma ^{-1}=\sqrt{1-|\mathbf{b}|^2/c^2}\end{equation}
and $P_{\mathbf{b}}$ is the orthogonal projection on the direction
of  $\mathbf{b}$ defined by
$P_{\mathbf{b}}\mathbf{r}=\frac{\langle \mathbf{b}|
\mathbf{r}\rangle }{|\mathbf{b}|^2}\mathbf{b}$.

 From this formula
 one derives the velocity addition as follows. Consider motion with
 uniform velocity $\mathbf{u}$ in system $K'$. The world line of
 this motion is $\left(
\begin{array}{c} t'\\ \mathbf{u}t'\end{array} \right).$ By use of
(\ref{lorentzvect}) this world line in system $K$ is
 \begin{equation} \gamma\left(
         \begin{array}{c}
              t'+  \frac{\mathbf{b}^T\mathbf{u}t'}{c^2} \\
              \mathbf{b} t'+ t'P_{\mathbf{b}}\mathbf{u}+\alpha t'(I- P_{\mathbf{b}})\mathbf{u}
          \end{array} \right)
 \end{equation}
 or
\begin{equation}\gamma t' \left(
         \begin{array}{c}
             1 +  \frac{\langle\mathbf{b}|\mathbf{u}\rangle}{c^2} \\
              \mathbf{b} +  \mathbf{u}_{\|}+\alpha \mathbf{u}_\bot
          \end{array} \right),\end{equation}
          where $\mathbf{u}_{\|}=P_{\mathbf{b}}\mathbf{u}$ denote
          the component of $\mathbf{u}$ parallel to $\mathbf{b}$ and
          $\mathbf{u}_\bot=(I-P_{\mathbf{b}})\mathbf{u}$ denote
          the component of $\mathbf{u}$ perpendicular  to $\mathbf{b}.$
This define a uniform motion in system $K$ with velocity, called
the relativistic velocity sum $\mathbf{b} \oplus \mathbf{u}.$
Thus, we get
\begin{equation}\label{veladd}
  \mathbf{b} \oplus \mathbf{u}=\frac{ \mathbf{b} +  \mathbf{u}_{\|}+\alpha \mathbf{u}_\bot}{1 +
   \frac{\langle\mathbf{b}|\mathbf{u}\rangle}{c^2}},
\end{equation}
which is the well-known Einstein velocity addition formula.

In case  $\mathbf{b}$ and  $ \mathbf{u}$ are parallel,
 this formula become:
\begin{equation}\label{veladdpar}
  \mathbf{b} \oplus \mathbf{u}=\frac{\mathbf{b} +
  \mathbf{u}}{1 +
   \frac{{b}{u}}{c^2}},
\end{equation}
and in case  $\mathbf{u}$ is  perpendicular to $ \mathbf{b}$ the
formula become:
\begin{equation}\label{veladdorth}
  \mathbf{b} \oplus \mathbf{u}=\mathbf{b} +\alpha(\mathbf{b})
  \mathbf{u}.
\end{equation}
Note that the velocity addition is commutative only for parallel
velocities.

We denote by $D_v$ the set of all relativistically admissible
velocities in an inertial frame $K$. This set is defined by
\begin{equation}\label{velballdef}
 D_v=\{\mathbf{v}: \; \mathbf{b}\in R^3, \;|\mathbf{b}|<c\}.
\end{equation}
The Lorentz transformation (\ref{lorentzvect}) acts on the
velocity ball $D_v$ as
\begin{equation}\label{boostvelball}
 \varphi _{\mathbf{b}}(\mathbf{v})=\mathbf{b} \oplus \mathbf{v}=
 \frac{ \mathbf{b} +  \mathbf{u}_{\|}+\alpha \mathbf{u}_\bot}{1 +
   \frac{\langle\mathbf{b}|\mathbf{v}\rangle}{c^2}},
\end{equation}
with $\alpha$ defined by (\ref{alpha}). It can be shown \cite{F04}
that the map $\varphi _{\mathbf{b}}$ is a projective (preserving
line segments) map of $D_v$.

We denote by $Aut _p(D_v)$ the group of all projective
automorphisms of the domain $D_v$. The map $\varphi _{\mathbf{b}}$
belongs to $Aut _p(D_v)$ and transforms any relativistically
admissible  velocity $\mathbf{v}\in D_v$ of the system $K'$, which
is moving parallel to $K$ with relative velocity $\mathbf{b}$, to
a corresponding unique velocity
$\varphi_{\mathbf{b}}(\mathbf{v})\in D_v$ in $K$.  Let $\psi$ be
any projective automorphism of $D_v$. Set $\mathbf{b}=\psi (0)$
and $U= \varphi_{\mathbf{b}}^{-1}\psi$. Then $U$ is an isometry
and represented by an orthogonal matrix. Thus, the group $Aut
_p(D_v)$ of all projective automorphisms is
\begin{equation}\label{AutD0}
 Aut _p(D_v)=\{\varphi_{\mathbf{b},U}=\varphi_{\mathbf{b}}U :
  \mathbf{b} \in D_v, \; U\in O(3)\}.
\end{equation}
This group represent the velocity transformation between arbitrary
two inertial systems and provide a representation of the Lorentz
group.

 Note that the Lorentz group representation defined by space-time
 transformations (\ref{lorentzvect}) between two inertial systems
is valid only if at time $t=0$ the origins of the two systems
coincide, while the velocity transformation (\ref{AutD0}) between
two inertial systems holds for arbitrary systems without any
limitation.

\section{Relativistic dynamics}

It is well known that a force generates a velocity change, or
acceleration. There are two types of forces. The first type
generates changes in the \textit{magnitude} of the velocity and
can be considered a velocity boost. An example is the force of an
electric field on a charged particle. The second type of force
generates a change in the \textit{direction} of the velocity - a
rotation or, equivalently, acceleration in a direction
perpendicular to the velocity of the object. An example is a
magnetic field acting on a moving charge. Thus a force can be
considered as a generator of velocity change. During the time
evolution, the velocity of an object cannot leave the velocity
ball $D_v$. Therefore, it is natural to assume that the generator
of a relativistic evolution \index{generator!relativistic
evolution} is an element of the Lie algebra $aut _p(D_v)$, which
consists of the generators of the group $Aut _p(D_v)$ generated by
velocity addition. We will call a relativistic motion generated by
a constant uniform force \textit{motion with uniform
acceleration}.

 To define the elements of $aut_p(D_v)$, consider
differentiable curves $g (s)$ from a neighborhood $I_0$ of $0$
into $Aut _p(D_v)$, with $g (0)=\varphi_{0,I}$, the identity of
$Aut _p(D_v)$. Any such $g(s)$ has the form
\begin{equation}\label{gamas}
g(s)=\varphi_{\mathbf{b}(s),U(s)},
\end{equation}
where $\mathbf{b}:I_0 \rightarrow D_v$ is a differentiable
function satisfying $\mathbf{b}(0)=\mathbf{0}$ and
$U(s):I_0\rightarrow O(3)$ is differentiable and satisfies
$U(0)=I$. We denote by $\delta$ the element of $aut_p(D_v)$
generated by $g (s)$. By direct calculation (see \cite{F04}) we
get
\begin{equation}\label{autD}
\delta (\mathbf{v})=\frac {d}{ds}g (s)(\mathbf{v})\Bigr|_{s=0}=
\mathbf{E}+A\mathbf{v}-c^{-2}\langle\mathbf{v}|\mathbf{E}\rangle\mathbf{v},
\end{equation}
where $\mathbf{E}=\mathbf{b}'(0)\in R^3$ and $A=U'(0)$ is $3\times
3$ skew-symmetric matrix. Defining $\mathbf{B}=\left(
         \begin{array}{c}
             a_{23} \\  -a_{13} \\  a_{12} \end{array} \right)$, we have
\begin{equation}\label{AandB}
  A\mathbf{v}=\mathbf{v}\times \mathbf{B},
\end{equation}
where $\times$ denotes the vector product in $R^3$. Thus, the
\index{Lie algebra!$aut_p(D_v)$} Lie algebra
\begin{equation}\label{autD1}
 aut_p(D_v)=\{\delta _{\mathbf{E},\mathbf{B}} :\mathbf{E},\mathbf{B} \in R^3  \},
\end{equation}
where $\delta _{\mathbf{E},\mathbf{B}}:D_v \rightarrow R^3$ is the
vector field defined by
\begin{equation}\label{deltaEB}
 \delta _{\mathbf{E},\mathbf{B}}(\mathbf{v})=
\mathbf{E}+\mathbf{v}\times
\mathbf{B}-c^{-2}\langle\mathbf{v}\,|\mathbf{E}\rangle\mathbf{v}.
\end{equation}
Note that any $\delta (\mathbf{v})$ is a polynomial in
$\mathbf{v}$ of degree less than or equal to 2. The elements of
$aut _p(D_v)$ transform between two inertial systems in the same
way as the electromagnetic field strength.

Evolution described by a relativistic dynamic equation must
preserve the ball $D_v$ of all relativistically admissible
velocities. If we consider the force as an element of $aut
_p(D_v)$, the equation of evolution of a charged particle with
charge $q$ and \textit{rest-mass} $m_0$  using the generator
$\delta_{\mathbf{E},\mathbf{B}}\in aut _p(D_v)$ is defined
by\begin{equation} \frac{d\mathbf{v}(\tau)}{d\tau
}=\frac{q}{m_0}\delta
_{\mathbf{E},\mathbf{B}}(\mathbf{v}(\tau)),\end{equation}  or
\begin{equation}\label{relevolution}
\frac{d\mathbf{v}(\tau)}{d\tau
}=\frac{q}{m_0}(\mathbf{E}+\mathbf{v}(\tau) \times
\mathbf{B}-c^{-2}\langle\mathbf{v}(\tau)|\mathbf{E}\rangle
\mathbf{v}(\tau)),
\end{equation}
 where $\tau$ is the proper time of the particle.
 It can be shown that this formula coincides with the
well-known formula
$$\frac{d(m\mathbf{v})}{dt }=q(\mathbf{E}+ \mathbf{v}\times \mathbf{B}).$$

Thus, the flow generated by an electromagnetic field is defined by
elements of the Lie algebra $aut_p(D_v)$, which are, in turn,
vector field polynomials in $\mathbf{v}$ of degree 2. The linear
term of this field comes from the magnetic force, while the
constant and the quadratic terms come from the electric field. If
the electromagnetic field $\mathbf{E},\mathbf{B}$ is constant, for
any given $\tau$ the solution of (\ref{relevolution}) is an
element $\varphi _{\mathbf{b}(\tau ),U(\tau)}\in Aut _p (D_v)$ and
the set of such elements form a one-parameter subgroup of $Aut _p
(D_v)$. This subgroup is a geodesic under the invariant metric on
the group. It can be shown by same argument (if we set
$\mathbf{B}=0$) that the dynamic equation of evolution in
relativistic \textit{mechanics} is also defined by elements of
$aut_p(D_v)$ .

Explicit solution of the evolution equation (\ref{relevolution})
exist only for constant electric $\mathbf{E}$ or constant magnetic
$\mathbf{B}$ fields. If both fields are present, even in case when
there is an invariant plane and the problem could be reduced to
one complex variable, there are no direct explicit solutions. The
reason to this is that equation  (\ref{relevolution}) is not
complex analytic. Complex analyticity is connected with conformal
maps, while the transformations on the velocity ball are
projective. All currently known explicit solutions
\cite{Baylis},\cite{Takeuchi02} and \cite{FS} use some
substitutions that in the new variable the transformations become
conformal.

\section{Explicit solutions for motion of a charge in constant,
uniform, and mutually perpendicular electric and magnetic fields}

\begin{figure}[h!]
  \centering
 \scalebox{0.5}{\includegraphics{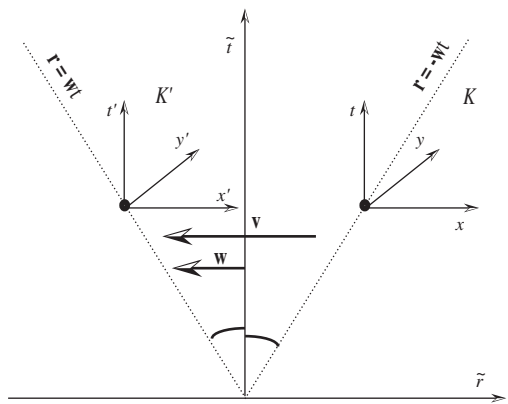}}
  \caption[The physical meaning of symmetric velocity.]
  { The physical meaning of symmetric velocity. Two inertial
  systems $K$ and $K'$ with relative velocity $\mathbf{v}$
between them are viewed from the system connected to their center.
In this system, $K$ and $K'$ are each moving with velocity $\pm
\mathbf{w}$. }\label{svelmeaning2}
\end{figure}

To obtain explicit solutions of the problem we associate with any
velocity $\mathbf{v}$ a new dynamic variable called the
\textit{symmetric velocity} $\mathbf{w_s}$. The symmetric velocity
$\mathbf{w_s}$ and its corresponding velocity $\mathbf{v}$ are
related by
\begin{equation}\label{velandsymvel}
 \mathbf{v}=\frac{\mathbf{w_s}+\mathbf{w_s}}{1+
 \frac{|\mathbf{w_s}|}{c}\,\frac{|\mathbf{w_s}|}{c}}=\frac{2\mathbf{w_s}}{1+|\mathbf{w_s}|^2/c^2}.
\end{equation}
The physical meaning of this velocity is explained in Figure
\ref{svelmeaning2}.

Instead of $\mathbf{w_s}$, we shall find it more convenient to use
the unit-free vector $\mathbf{w}=\mathbf{w_s}/c$, which we call
the \textit{s-velocity}. The relation of a velocity $\mathbf{v}$
to its corresponding s-velocity is
\begin{equation}\label{velandsvel}
\mathbf{v}=\Phi(\mathbf{w})=\frac{2c\mathbf{w}}{1+|\mathbf{w}|^2},
\end{equation}
where $\Phi$ denotes the function mapping the s-velocity
$\mathbf{w}$ to its corresponding velocity $\mathbf{v}$. The
s-velocity has some interesting and useful mathematical
properties. The set of all three-dimensional relativistically
admissible s-velocities forms a unit ball
\begin{equation}
D_s = \{\mathbf{w}\in R^3:\; |\mathbf{w}|<1\}.
\end{equation}

Corresponding to the Einstein velocity addition equation, we may
define an addition of s-velocities in $D_s$ such that
\begin{equation}\label{einsteinaddsymaddition}
 \Phi ( \mathbf{b} \oplus _s \mathbf{w})= \Phi(\mathbf{b}) \oplus
  _E  \Phi (\mathbf{w}).
\end{equation}
A straightforward calculation leads to the corresponding equation
for s-velocity addition:
\begin{equation}\label{symveladdition}
  \mathbf{b} \oplus _s \mathbf{w}=
  \frac{(    1+|\mathbf{w}|^2  +2<\mathbf{b}\mid \mathbf{w}> )\mathbf{b}+
             (1-|\mathbf{b}|^2 )\mathbf{w}
        }
        {    1+|\mathbf{b}|^2|\mathbf{w}|^2 +
             2<\mathbf{b}  \mid \mathbf{w}>
        }.
\end{equation}

Equation (\ref{symveladdition}) can be put into a more convenient
form if, for any $\mathbf{b}\in D_s$, we define a map
$\Psi_{\mathbf{b}}:D_s\to D_s$ by
\begin{equation}\label{psia}
 \psi_{\mathbf{b}}(\mathbf{w})\equiv\mathbf{b} \oplus _s
 \mathbf{w}.
\end{equation}
This map is an extension to $D_s \in R^n$ of the M\"{o}bius
addition on the complex unit disc. It defines a \textit{conformal}
map on $D_s$. The motion of a charge in $\mathbf{E} \times
\mathbf{B}$ fields is two-dimensional if the charge starts in the
plane perpendicular to $\mathbf{B}$, and in this case
Eq.(\ref{symveladdition}) for s-velocity addition is somewhat
simpler.  By introducing a complex structure on the plane $\Pi$,
which is perpendicular to $\mathbf{B}$, the disk
$\Delta=D_s\cap\Pi$ can be identified as a unit disc $|z|<1$
called the Poincar\'{e} disc. In this case the s-velocity addition
defined by Eq.(\ref{symveladdition}) becomes
\begin{equation}\label{mobius}
 a \oplus _s w=\psi_{{a}}({w})=\frac{a+w}{1+\overline{a}w},
\end{equation}
which is the well-known M\"{o}bius transformation of the unit
disk.

By using the $s$ velocity we can rewrite \cite{F04},\cite{FS} the
relativistic Lorentz force equation $$\frac{d}{dt}(\gamma
m\mathbf{v})=q({\bf E} +\mathbf{v}\times {\bf  B})$$ as
\begin{equation}\label{EB18}
\frac{m_0 c}{q}\,\frac{d\mathbf{w}}{d\tau}
=\left(\frac{1+|\mathbf{w}|^2}{2}\right){\bf{E}}+c\mathbf{w}\times
{\bf{B}}-\mathbf{w}<\mathbf{w}|{\bf{E}}>,
\end{equation}
which is the relativistic Lorentz force equation for the
s-velocity $\mathbf{w}$ as a function of the proper time $\tau$.

We now use Eq.(\ref{EB18}) to find the s-velocity of a charge $q$
in uniform, constant, and mutually perpendicular electric and
magnetic fields.  Since all of the terms on the right hand side of
Eq. (\ref{EB18}) are in the plane perpendicular to ${\mathbf{B}}$,
if $\mathbf{w}\in\Pi$, therefore $d{\mathbf{w}}/{d\tau}$ is in the
plane $\Pi$ perpendicular to ${\mathbf{B}}$.  Consequently, if the
initial s-velocity is in the plane perpendicular to
${\mathbf{B}}$, ${\mathbf{w}}$ will remain in the this plane and
the motion will be two dimensional.

Working in Cartesian coordinates, we choose
\begin{equation}\label{defEB}
\mathbf{E}=(0,E,0),\; \mathbf{B}=(0,0,B),\; \mbox{and}\;
\mathbf{w}=(w_1,w_2,0).
\end{equation}
By introducing a complex structure in $\Pi$ by denoting  $w=w_1 +
iw_2$ the evolution equation Eq.(\ref{EB18}) get the following
simple form:
\begin{equation}\label{EBmain} \frac{dw}{d\tau}
=i\Omega\left(w^2-2\widetilde{B}w+1\right),
\end{equation}
where
\begin{equation}\label{OmegaBtilde}
\Omega\equiv\frac{qE}{2m_0
c}\;\;\mbox{and}\;\;\widetilde{B}\equiv\frac{cB}{E}.
\end{equation}
The solution of Eq.(\ref{EBmain}) is unique for a given initial
condition
\begin{equation}\label{initialconditionEB}
 w(0)=w_0,
\end{equation}
where the complex number $w_0$ represents the initial s-velocity
$\mathbf{w}_0=\Phi^{-1}(\mathbf{v}_0)$ of the charge.

Integrating Eq.(\ref{EBmain}) produces the equation
\begin{equation}\label{intEB2}
\int\frac{dw}{w^2-2\widetilde{B}w+1} =i\Omega \tau +C,
\end{equation}
where the constant $C$ is determined from  the initial condition
(\ref{initialconditionEB}). The way we evaluate this integral
depends upon the sign of the discriminant $4\widetilde{B}^2-4$
associated with the denominator of the integrand. If we define
\begin{equation}\label{discriminantEB}
\Delta\equiv \widetilde{B}^2-1=\frac{(cB)^2-E^2}{E^2},
\end{equation}
then the three cases $E<cB, E=cB\; \mbox{and}\; E>cB$ correspond
to the cases $\Delta$ greater than zero, equal to zero, and less
than zero.

\noindent {\bf Case 1} Consider first the
case\begin{equation}\Delta=
((cB)^2-{E}^2)/{E}^2)>0\Longleftrightarrow
{E}<cB\;\mbox{and}\;\widetilde{B}>1. \end{equation} The
denominator of the integrand in (\ref{intEB2}) can be rewritten
as\begin{equation} w^2-2\widetilde{B}w+1=(w-\alpha _1)(w-\alpha
_2), \end{equation} where $\alpha _1$ and $\alpha _2$ are the
real, positive roots
\begin{equation}\label{EBroots}
\alpha _1=\widetilde{B}-\sqrt{\widetilde{B}^2-1}\;\;\mbox{and}\;\;
\alpha _2=\widetilde{B}+\sqrt{\widetilde{B}^2-1}.
\end{equation}
and the solution become:
\begin{equation}\label{solevoleqncase3}
w(\tau)=\frac{\alpha _1+Ce^{-i\nu\tau}} {1+\alpha
_1Ce^{-i\nu\tau}}=\alpha _1\oplus _s Ce^{-i\nu\tau},
\end{equation}
where $\nu=2\sqrt{\Delta}\Omega$. This equation shows that in a
system K' moving with s-velocity $\alpha _1$ relative to the lab,
the s-velocity of the charge corresponds to circular motion with
initial s-velocity
\begin{equation}\label{initialC}
 C=\psi _{-\alpha _1}(w_0).
\end{equation}

From Eqs.(\ref{velandsvel}) and (\ref{EBroots}) it follows that
the lab velocity corresponding to s-velocity $\alpha _1$ is
\begin{equation}\label{driftveldef}
  \frac{2c\alpha _1}{1+|\alpha _1|^2}=(E/B)\mathbf{i}=\mathbf{v}_d=v_d\mathbf{i},
\end{equation}
which is the well-known $\mathbf{E}\times \mathbf{B}$ drift
velocity. Applying the map $\Phi$ defined in
Eq.(\ref{einsteinaddsymaddition}) to both sides of
(\ref{solevoleqncase3}), we get
\begin{equation}\label{solevoleqncase3a}
\mathbf{v}(\tau)=\mathbf{v}_d\oplus _E e^{-i\nu\tau}\Phi (C).
\end{equation}
Eq.(\ref{solevoleqncase3a}) says that the total velocity of the
charge,  as a function of the proper time, is the sum of a
constant drift velocity $\mathbf{v}_d= (E/B) \mathbf{i}$ and
circular motion, as expected.

If $\mathbf{v}(0)=0$, then the $s$ velocity of the charge as a
function of the proper time is ${\mathbf{b}}(\tau)=\alpha
_1(1-\cos\nu\tau,-\sin\nu\tau),$ the velocity of the charge is
\[\mathbf{v}_{\mathbf{b}}(\tau)={v}
_d(1-\cos\nu\tau,-\sin\nu\tau),\] the position of the charge is
\begin{equation}\label{posproptime}
\mathbf{r}_{\mathbf{b}}(\tau)=\int_0^{\tau}\gamma\mathbf{v}(\tau')\,d\tau'
= \frac{\gamma_dv_d}{\nu}\left(\gamma_d
(\nu\tau-\sin{\nu\tau}),(\cos\nu\tau-1)\right)\end{equation} and
the lab time $t$  as a function of the proper time is
\begin{equation}\label{tastaucase3}
 t_{\mathbf{b}}(\tau)=\int_0^{\tau}\gamma(\tau')d
 \tau'=\frac{\gamma_d^2}{\nu}\left(\nu\tau-\frac{v_d^2}{c^2}\sin{\nu\tau}\right),
\end{equation}
where $\gamma_d=\gamma(\mathbf{v}_d)$. 

\noindent {\bf Case 2}\, Next consider the case $\Delta=
((cB)^2-{E}^2)/{E}^2=0\Longleftrightarrow
{E}=c{B}\;\mbox{and}\;\widetilde{B}=1$. The denominator in the
integrand of (\ref{intEB2}) is $w^2-2w+1=(w-1)^2$ and its solution
is
\begin{equation}\label{soleveqncase2}
w(\tau)=1-\frac{1}{i\Omega\tau +C}
\end{equation}
with $ C=-\frac{1}{w_0-1}.$ 

If the initial velocity is zero, $C=1$ the $s$ velocity of the
charge as a function of the proper time is
\[{\mathbf{b}}(\tau)=\frac{(\Omega^2\tau^2,\Omega\tau)}{1+\Omega^2\tau^2},\]
the velocity of the charge
\[\mathbf{v}_{\mathbf{b}}(\tau)=\frac{2c(\Omega^2\tau^2,\Omega\tau)}{1+2\Omega^2\tau^2},\]
the position of the charge is
\begin{equation}\label{proppos2}
\mathbf{r}_{\mathbf{b}}(\tau)=2c\left(\frac{\Omega^2\tau^3}{3},\frac{\Omega\tau^2}{2}\right)
\end{equation} and the lab time as a function of the proper time
is
\begin{equation}\label{timescase2}
t_{\mathbf{b}}(\tau)=\int_0^\tau \gamma d\tau=\tau+\frac{2\Omega
^2}{3}\tau ^3.
\end{equation}
Equations  (\ref{proppos2})  and (\ref{timescase2})  give the
complete solution for this case. 

\noindent {\bf Case 3}\, Consider the case $\Delta= ((cB)^2-
E^2)/{E}^2<0\;\Longleftrightarrow
{E}>cB\;\mbox{or}\;\widetilde{B}<1$.

Just as in Case 1, we rewrite the denominator of the integrand in
Eq. (\ref{intEB2}) as $w^2-2\widetilde{B}w+1=(w-\alpha
_1)(w-\alpha _2),$ where\begin{equation}\alpha
_1=\widetilde{B}-i\delta\;\;\mbox{and}\;\; \alpha
_2=\widetilde{B}+i\delta=\overline{\alpha}_1\end{equation}   and
$\delta=\sqrt{1-\widetilde{B}^2}>0.$ By introducing
\begin{equation}
\nu=\left(\frac{q}{mc}\right)\sqrt{E^2-(cB)^2}.
\end{equation}
and an s-velocity ${w}_d
\equiv{\widetilde{B}}/\left({1+\delta}\right)$ we can write the
solution as:
\begin{equation}\label{solevoleqncase1}
w(\tau)=w _d\oplus _s (i\tanh(\nu\tau)\oplus _s \widetilde{w}_0),
\end{equation}
where $\widetilde{w}_0=\psi_{-w_d}(w_0).$ 

For the velocity of the charge we get
\begin{equation}\label{solevoleqncase1a}
\mathbf{v}(\tau)=\mathbf{v}_d\oplus _E
(c\tanh(2\nu\tau)\mathbf{j}\oplus  \widetilde{\mathbf{v}}_0 ),
\end{equation}
where $\mathbf{v}_d =(c^2 B/E)\mathbf{i}$ is the drift velocity
and $\widetilde{\mathbf{v}}_0$ is the initial velocity in the
drift frame. From this it follows that for initial zero velocity,
the velocity of the charge as a function of the proper time is
\[\mathbf{v}_{\mathbf{b}}(\tau)= \frac{\left(\gamma_d {v} _d(\cosh(\nu'\tau)-1)
,c\sinh(\nu'\tau)\right)}{\gamma_d
\left(\cosh(\nu'\tau)-{v_d^2}/{c^2}\right)},
\] its position
\begin{equation}\label{finalr32}
\mathbf{r}_{\mathbf{b}}(\tau)=\int_0^{\tau}\gamma\mathbf{v}(\tau')\,d\tau'
= \frac{\gamma_d}{\nu'}\left(\gamma_d v_d
(\sinh(\nu'\tau)-\nu'\tau),c(\cosh(\nu'\tau)-1)\right)\end{equation}
and the lab time $t$ as a function of the proper time is
\begin{equation}\label{finalt3}
 t_{\mathbf{b}}(\tau)=\int_0^{\tau}\gamma(\tau')d \tau'=\gamma_d^2\left(\frac{\sinh(\nu'\tau)}{\nu'}-\frac{v_d^2}{c^2}\tau\right).
\end{equation}
Equations (\ref{finalr32}) and (\ref{finalt3}) together give the
complete solution for this case.

In all cases, for arbitrary initial velocity $\mathbf{v}_0,$ the
velocity of the charge at its proper time $\tau$ will be given by
(see \cite{F04} p.87)
\begin{equation}\label{veltransfundeEB}
  \mathbf{v}(\tau)=\varphi_{\mathbf{v}_{\mathbf{b}}(\tau)}(U_{\mathbf{b}}(\tau)\mathbf{v}_0),
\end{equation}
where $\varphi$ is defined by (\ref{boostvelball}),
$\mathbf{v}_{\mathbf{b}}$ was defined in each case separately and
$U_{\mathbf{b}}(\tau)$ is a rotation in the plane $x,y$ given by
the complex number
\begin{equation}\label{U_b}
  \frac{1-\widetilde{B}\mathbf{b}(\tau)}{1-\widetilde{B}\overline{\mathbf{b}}(\tau)}.
\end{equation}
This mean that for the electromagnetic field
$\mathbf{E},\mathbf{B}$ in an inertial system $K$, with
$\mathbf{E}$ perpendicular to $\mathbf{B}$, in the frame
$\tilde{K}$ boosted with $\mathbf{v}_{\mathbf{b}}(\tau)$ and
rotated with $U_{\mathbf{b}}(\tau)$ with respect to our inertial
system all charged particles of the same mass and charge will
continue to move with their initial velocity. This mean that with
respect to system $\tilde{K}$ the motion of charged particles is
not affected by the electromagnetic field. This is similar to
Equivalence Principle for the gravitational field.

\section{Space-time transformations from an inertial system to a uniformly accelerated
and rotating system assuming the clock-hypothesis}

Results of the previous section could be applied to find the
space-time transformations between an inertial system $K$ and a
uniformly accelerated and rotating system $\tilde{K}$. The motion
of a charge in a constant uniform electric field is considered as
a uniformly accelerated motion, while the motion in a magnetic
field is considered as constant rotation. Thus, constant uniform
acceleration and rotation may be described by the action of a
constant electromagnetic field $\mathbf{E},\mathbf{B}$ on charged
particles. In this section we will deal only with uniformly
accelerated and rotating motion for which the axes of rotation is
perpendicular to direction of the acceleration. This corresponds
to perpendicularity of the corresponding electric and magnetic
fields.

Let $K$ denote an inertial system with with origin $O$
and space-time coordinates $\left( \begin{array}{c} t\\
\mathbf{r}\end{array} \right).$  Let a system $\tilde{K}$ with
origin $\tilde{O}$ move with constant uniform acceleration and
rotation is described by the action of a constant electromagnetic
field $\mathbf{E},\mathbf{B}$, with $\mathbf{E}$ perpendicular to
$\mathbf{B}$, on charged particles. Without loss of generality we
may assume that $\mathbf{E},$ generating the acceleration, is in
the direction of the $y$-axis, \textit{i.e.} $\mathbf{E}=(0,E,0)$
and $\mathbf{B},$ generating the rotation around the $z$-axis,
\textit{i.e.} $\mathbf{B}=(0,0,B)$. We assume that the clocks and
the space axes at time $t=0$ in systems $K$ and $\tilde{K}$ were
synchronized.  A charge positioned at common origin of the systems
at time $t=0$ remains at $\tilde{O}$ for any time $t>0$. Thus, by
the results of the previous section, the world line of this charge
or of $\tilde{O}$ is $\left(
\begin{array}{c} t_{\mathbf{b}}(\tau)\\ \mathbf{r}_{\mathbf{b}}(\tau)\end{array} \right),$
where $\tau$ denote the proper time of the charge and
$t_{\mathbf{b}}(\tau)$ and $\mathbf{r}_{\mathbf{b}}(\tau)$ are
defined by the appropriate formulas for each of the 3 cases.

For a given time $t_0$ we denote by $K'$ an inertial system with
origin $O'$ which is positioned and have the same velocity as
$\tilde{O}$ at time $t_0$ and have common space axes at time $t_0$
with system $\tilde{K}$. The system $K'$ is called a
\textit{comoving system} to system $\tilde{K}$ at time $t_0$. The
\textit{Clock hypothesis} state that the time in $\tilde{K}$ is
the same as the time in $K'$. As we will see later, the Clock
hypothesis imply that the space coordinates in $\tilde{K}$ and
$K'$ are the same. Thus, the space-time transformations from $K$
to $\tilde{K}$ coincide with the transformations between $K$ and
$K'.$

Consider an event that occurs  at $\left(
\begin{array}{c} \tilde{t}\\ \mathbf{\tilde{r}}\end{array} \right)$
in the uniformly accelerated and rotated system $\tilde{K}$.Let
$K'$ be the inertial system comoving to $\tilde{K}$ at time
$t_0=\tilde{t}$. By the Clock hypothesis this event has the same
space-time coordinates in $K'$. The position of the origins $O'$
and $\tilde{O}$ is $\mathbf{r}_{\mathbf{b}}(\tilde{t})$ in $K$ and
the time at the clock at $\tilde{O}$ shows time $\tilde{t}$ which
correspond to $t_{\mathbf{b}}(\tilde{t})$ in system $K$. Moreover
the space axes in $K$ are rotated with respect to the space axes
in $K'$ by $U_{\mathbf{b}}(\tilde{t}),$ defined by (\ref{U_b}).
The relative velocity of $K'$ and $K$ is given by
$\mathbf{v}_{\mathbf{b}}(\tilde{t}),$ defined by the appropriate
formulas for each of the 3 cases. We use a modification of the
space-time Lorentz transformations (\ref{lorentzvect0}) between
two inertial systems which were not synchronized at time $t=0$ for
the space-time transformation from $K'$ to $K$. This imply that
the space-time coordinates of the considered event in $K$ are
\begin{equation}\label{spactimeEBcomovig}
  \left(\begin{array}{c} {t}\\ \mathbf{{r}}\end{array} \right)=
  \left(\begin{array}{c} t_{\mathbf{b}}(\tilde{t})\\ \mathbf{r}_{\mathbf{b}}(\tilde{t})\end{array}
  \right)+\left(\begin{array}{c} \gamma c^{-2}\langle
  \mathbf{v}_{\mathbf{b}}(\tilde{t})| U_{\mathbf{b}}(\tilde{t})\mathbf{\tilde{r}}\rangle
  \\(\gamma P_{\mathbf{v}_{\mathbf{b}}(\tilde{t})}+(I-P_{\mathbf{v}_{\mathbf{b}}(\tilde{t})}))
   U_{\mathbf{b}}(\tilde{t})\mathbf{\tilde{r}}\end{array} \right),
\end{equation}
where $\gamma=\gamma (\mathbf{v}_{\mathbf{b}}(\tilde{t})).$

If we do not assume the validity of the Clock Hypothesis, the
above transformations will hold with respect to the comoving frame
$K'$ only. Thus, in order to describe general space-time
transformations between two systems uniformly accelerated with
respect to the same inertial system, it is enough to describe such
transformations only for systems which at some initial time have
the same velocity, which were called {comoving} systems. For
example, systems $K',$ which is an inertial system, could be
considered as moving with uniform zero acceleration with respect
to the inertial system $K$ and system $\tilde{K}$, which is
uniformly accelerated with respect to the inertial system $K$ are
comoving with respect to $t_0$. Thus, they provide an example of
uniformly accelerated comoving systems. Such transformation are
described in \cite{FG5}.


\begin{thebibliography}{9}
%
\bibitem{Baylis} W. E. Baylis,
\textit{ Electrodynamics, A Modern Geometric Approach}, Progress
in Physics \textbf{17}, Birkh\"{a}user, Boston, 1999.
%

\bibitem{B65}  M. Born, \textit{Einstein's Theory of Relativity }, Dover
Publications, New York, 1965.

%
\bibitem{Chen} C. X. Chen, G. Papini, N. Mobed, G. Lambiaise, G.  Scarpetta,
 Nuovo Cimento 114B (1999) 1335.




\bibitem{F04}Y. Friedman, \textit{ Homogeneous balls and their Physical
Applications}, Progress in Mathematical Physics 40,
(Birkh\"{a}user, Boston, 2004.

\bibitem{FG02/1} Y. Friedman, Yu. Gofman, Why Does the Geometric
Product Simplify the Equations of Physics?, \textit{International
Journal of Theoretical Physics,} \textbf{41} (2002), 1861--1875.

\bibitem{FG02/2} Y. Friedman, Yu. Gofman, Relativistic Linear
Spacetime Transformations Based on Symmetry, \textit{Foundations
of Physics,} \textbf{32} (2002), 1717--1736.

 \bibitem{FG4}Y.  Friedman, Yu. Gofman, in Gravitation, Cosmology and Relativistic
 Astrophysics, A Collection of Papers (Kharkov National
University,  Kharkov, Ukraine, 2004), 102.

\bibitem{FG5}Y.  Friedman, Yu. Gofman, Kinematic relations between
relativistically accelerated systems in flat space-time based on
symmetry, arxiv/gr-qc/0509004.

 \bibitem{FS} Y. Friedman, M.Semon,
  Relativistic acceleration of charged particles in
 uniform and mutually perpendicular electric and magnetic fields as viewed in
the laboratory frame, \textit{
 Phys. Rev.} E \textbf{72} (2005), 026603.

%
%
\bibitem{Lambiaise} G. Lambiaise,  G. Papini, G.  Scarpetta, Nuovo
Cimnto 114 (1999) 189.

%
%

%
%
%
%

\bibitem{P58}  W. Pauli, Theory of Relativity (Pergamon
Press, London, 1958).


\bibitem{shuller02} F. P. Schuller, Ann. Phys. 299 (2002) 174.


\bibitem{scarpetta84} G. Scarpetta, Lett. Nuovo Cimento 41 (1984) 51.

\bibitem{Takeuchi02} S. Takeuchi, Phys. Rev. E\textbf{66}, 37402-1 (2002).

 \end{thebibliography}
\end{document}